\documentclass[11pt]{article}

\usepackage{acl}
\usepackage{booktabs}
\usepackage{tabularx}
\usepackage{array}
\usepackage{enumitem}
\usepackage{xcolor}
\usepackage{amsmath}
\usepackage{amssymb}
\usepackage{graphicx}
\usepackage{float}
\usepackage{microtype}
\usepackage{tikz}
\usepackage{pgfplots}

\pgfplotsset{compat=1.18}
\usetikzlibrary{arrows.meta,positioning,calc,shapes.geometric}
\newcommand{\method}{Squeez}

\title{\method: Task-Conditioned Tool-Output Pruning for Coding Agents}

\author{Ádám Kovács \\
  KR Labs \\
  \texttt{kovacs@krlabs.eu} \\
  \url{https://github.com/KRLabsOrg/squeez}}

\begin{document}
\maketitle

\begin{center}
\includegraphics[width=0.14\linewidth]{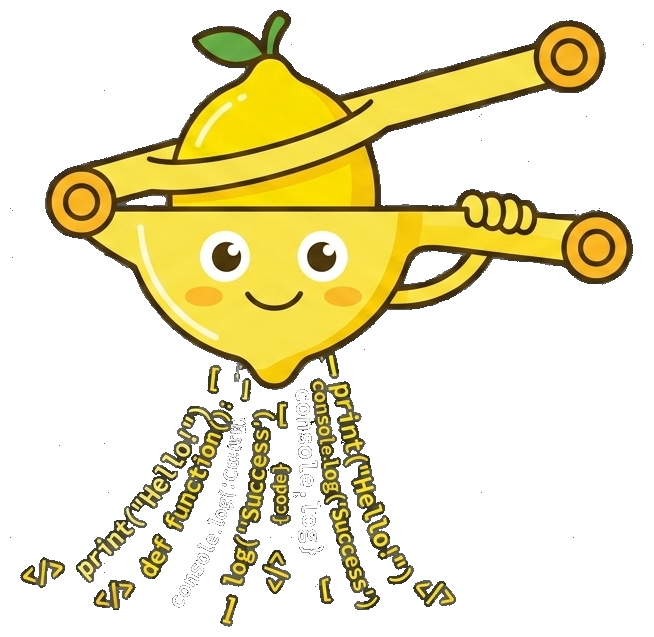}
\end{center}
\vspace{-0.6em}

\begin{abstract}
Coding agents repeatedly consume long tool observations even though only a
small fraction of each observation matters for the next step. We study
\emph{task-conditioned tool-output pruning}: given a focused query and one tool
output, return the smallest verbatim evidence block the agent should inspect
next. We introduce a benchmark of 11{,}477 examples built from SWE-bench
repository interactions and synthetic multi-ecosystem tool outputs, with a
manually curated 618-example test set. We fine-tune Qwen 3.5 2B with LoRA and
compare it against larger zero-shot models and heuristic pruning baselines. Our
model reaches 0.86 recall and 0.80 F$_1$ while removing 92\% of input
tokens, outperforming zero-shot Qwen 3.5 35B A3B by 11 recall points and all
heuristic baselines by a wide margin.
\end{abstract}

\section{Introduction}
Coding agents work over streams of file reads, grep hits, stack traces, build
logs, API responses, and version-control history
\citep{yang2024sweagent,wang2024opendevin}. In such observations, only a
small fraction of the tokens is relevant to the next decision, yet the agent
often has to re-read the entire output. This makes pruning a central efficiency
problem for agent systems
\citep{jiang2023llmlingua,jiang2024longllmlingua,hwang2024exit,
focusagent2025,ayanami2026swepruner}.

We study a narrow but practical version of this problem: given a focused query
and a single tool observation, return the smallest verbatim evidence block the
agent should inspect next. The task is not to solve the issue from one
observation, but to preserve the relevant evidence and discard the rest. This
distinguishes our setting from generic prompt compression and from
document-level retrieval compression, which typically operate on prose
documents or retrieved passages rather than mixed-format tool output.

We make three contributions. We formulate task-conditioned tool-output
pruning as a benchmark task for coding agents, release a dataset spanning real
SWE-bench repository interactions and synthetic multi-ecosystem tool outputs,
and show that a compact Qwen 3.5 2B model fine-tuned for this task
substantially outperforms larger zero-shot models and simple heuristic
baselines. Because the model can be served through vLLM or used as a CLI
filter on piped tool output, it can be inserted into existing coding-agent
stacks such as \href{https://openai.com/codex/}{Codex} or
\href{https://docs.anthropic.com/en/docs/claude-code/overview}{Claude Code}
with minimal changes to the surrounding agent loop. The
\href{https://huggingface.co/KRLabsOrg/squeez-2b}{model},
\href{https://huggingface.co/datasets/KRLabsOrg/tool-output-extraction-swebench}{dataset},
and \href{https://github.com/KRLabsOrg/squeez}{evaluation code and CLI}
are publicly available under the Apache 2.0 license.

\section{Related Work}
Prompt compression methods such as LLMLingua \citep{jiang2023llmlingua} and
LongLLMLingua \citep{jiang2024longllmlingua} compress long prompts before
generation. LLMLingua uses a coarse-to-fine compression pipeline with a budget
controller and iterative token removal, while LongLLMLingua adapts compression
to long-context settings with query-aware compression and document reordering.
These methods target prompt efficiency at the token or prompt-block level 
rather than query-conditioned extraction from a single mixed-format tool
observation. Abstractive summarization addresses a different problem again,
since it rewrites salient content rather than preserving verbatim source
evidence \citep{see2017get,liu2019text}.

Document-level context pruners are closer to our work. EXIT
\citep{hwang2024exit} performs extractive compression for
retrieval-augmented generation by selecting relevant portions of retrieved
textual context for a downstream question-answering stage. Provence
\citep{chirkova2025provence} formulates context
pruning as sequence labeling, combining pruning with reranking, and training on
diverse QA domains to obtain robust out-of-the-box RAG pruning. Open-source models such as Zilliz Semantic Highlight \citep{zilliz2025semantic}
adapt this idea to lightweight semantic highlighting, producing token- and sentence-level
relevance scores over retrieved documents for real-time highlighting. These
systems, however, assume retrieved passages or document text. They do not
target the mixed observations that arise in coding agents, where code, logs,
shell traces, metadata, and structured outputs are interleaved in a single
artifact.

Our setting is also related to agent-specific pruning. FocusAgent
\citep{focusagent2025} reduces the context available to web agents by trimming
web observations, while SWE-Pruner \citep{ayanami2026swepruner} targets coding
agents more directly by pruning repository code context. The nearest of these
to our work is SWE-Pruner, but its emphasis is repository-level code context
rather than single tool observations spanning files, logs, build outputs, and
other non-code modalities.

Finally, our task is related to extractive QA and supporting-evidence
supervision, where outputs are grounded in explicit spans or sentence sets
rather than free generation \citep{rajpurkar2016squad,thorne2018fever,
yang2018hotpotqa}. Verbatim evidence extraction has also been applied in
domain-specific QA, where lightweight extractive pipelines select grounded
sentences to prevent hallucination \citep{kovacs2025verbatim}. The common
concern is faithfulness: outputs should remain traceable to source evidence.
Our setting differs in focusing on query-conditioned pruning of tool
observations rather than answer production from prose documents.

\begin{figure}[t]
\centering
\begin{tikzpicture}[
  font=\footnotesize,
  >=Latex,
  box/.style={draw=black, rounded corners=3pt, align=left, inner sep=6pt,
    fill=gray!5, text width=0.84\columnwidth},
  arrow/.style={->, thick, shorten >=2pt, shorten <=2pt}
]
\node[box] (input) {
  \textbf{Input}\\[3pt]
  \emph{Query:} ``Find the traceback\\
  that explains the ImportError.''\\[3pt]
  \emph{Tool output:} 501 lines from\\
  \texttt{read\_file}, \texttt{grep},\\
  \texttt{pytest}, \texttt{git\_log}, \ldots
};

\node[box, below=0.8cm of input] (canon) {
  \textbf{Grounded spans}\\[3pt]
  \texttt{{\color{gray}183:} ):}\\
  \colorbox{yellow!25}{\texttt{184: super().\_\_init\_\_(}}\\
  \colorbox{yellow!25}{\texttt{185: \ \ import\_name=...}}\\
  \colorbox{yellow!25}{\texttt{191: self.name = name}}\\
  \texttt{{\color{gray}193:} self.subdomain = ...}\\[3pt]
  Gold: \texttt{(start\_line, end\_line)}
};

\node[box, below=0.8cm of canon] (output) {
  \textbf{Generative output}\\[3pt]
  \texttt{<relevant\_lines>}\\
  \texttt{184: super().\_\_init\_\_(}\\
  \texttt{185: \ \ import\_name=...}\\
  \texttt{191: self.name = name}\\
  \texttt{</relevant\_lines>}\\[3pt]
  Eval: recall, F$_1$, compression
};

\draw[arrow] (input) -- node[right, font=\scriptsize] {annotate} (canon);
\draw[arrow] (canon) -- node[right, font=\scriptsize] {derive} (output);

\end{tikzpicture}
\caption{System overview. The benchmark stores grounded spans over raw tool
output, then derives a generative training target. At evaluation, predicted
lines are compared to gold spans using recall, F$_1$, and compression.}
\label{fig:overview}
\end{figure}
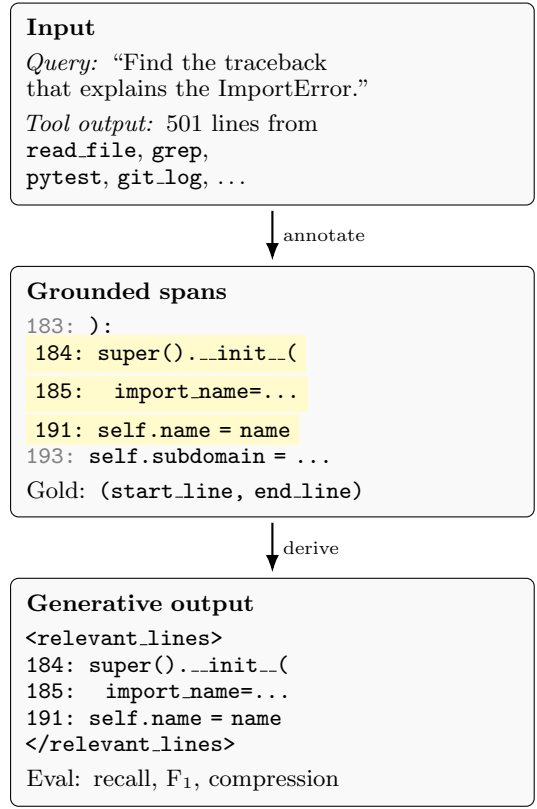

\section{Task Definition}
Figure~\ref{fig:overview} illustrates the overall pipeline.
The benchmark input is a pair $(q, o)$, where $q$ is a short task-conditioned
extraction query and $o$ is one raw tool observation. The output is a set of
one or more contiguous spans over $o$:
\[
Y = \{(s_1, e_1), \dots, (s_k, e_k)\},
\]
where each span refers to line indices in the original output.

The task is intentionally narrower than bug solving. The model is not asked to
infer the correct patch from one observation, but to extract the minimal
evidence block that would help an agent on the next reasoning step. Queries are
short, concrete, and tool-aware. They may be easy, medium, or moderately
semantic; they need not avoid lexical cues if those cues reflect the underlying
agent need. The key requirement is that they define a plausible pruning
decision rather than a full diagnosis.

\section{Dataset}
\subsection{Data Sources}
The benchmark draws on two sources of raw tool output.

\paragraph{SWE-bench derived.}
We clone repository snapshots from SWE-bench
\citep{jimenez2024swebench} and execute 14 tool types against them ---
file reads, grep, Git log and blame, test runners, linters, type checkers,
pip install, curl, and others --- collecting 10{,}713 raw observations that
reflect the kind of output a coding agent encounters during issue resolution.

\paragraph{Synthetic multi-ecosystem.}
To extend coverage beyond SWE-bench's Python-heavy distribution, we use
\texttt{openai/gpt-oss-120b} to generate 2{,}039 realistic tool outputs for
representative tasks in TypeScript, Go, Rust, Java, Docker, Terraform,
Kubernetes, and related build or deployment workflows. We also
construct 575 explicit negative examples by pairing mismatched queries and
tool outputs, where the correct pruning decision is to return nothing.

The released benchmark contains 11{,}477 examples in total:
9{,}205 SWE-derived, 1{,}697 synthetic positives, and 575 synthetic
negatives, covering 27 tool types;
Table~\ref{tab:tool-counts} lists the tool-family counts.

\subsection{Example Construction}
Each example pairs one focused query with one raw tool observation. Both
sources go through the same two-stage teacher-labeling pipeline using
\texttt{openai/gpt-oss-120b}. Given the background task and the raw tool
output, the teacher first writes a focused, tool-aware extraction query:
not the full issue statement, but a narrower local information need such as
finding the failure block, the relevant code region, or the commit entry that
matters for the next debugging step. It then selects the smallest contiguous
span or set of spans that answers that query.

The teacher sees a numbered rendering of the tool output for stable span
selection, but the released labels are always mapped back onto the original
raw text, so every target is a verbatim subset of the source. Positive
examples whose query cannot be supported by the observation are discarded.
For generative training, gold spans are linearized as XML-wrapped extracted
text, while the released benchmark stores the span coordinates over the
original tool output.

\subsection{Curation and Statistics}
We split SWE-derived examples by repository and synthetic examples by tool
family, yielding 10{,}508 training examples, 240 development examples, and 618
test examples. The held-out set was manually curated: from 729 candidate test
examples, 111 (15.2\%) were excluded as near-duplicates, trivial 1--2 line
outputs, overly broad spans, or incorrect annotations. The final released test
set therefore contains only manually reviewed examples.

Table~\ref{tab:source-stats} summarizes the source composition of the released
benchmark. Table~\ref{tab:tool-stats} shows the largest tool families. The
distribution is intentionally heterogeneous: the benchmark mixes very short
observations such as Python exceptions and test summaries with long file reads,
type-check outputs, and container logs. This variation is one reason simple
truncation and lexical retrieval are weak baselines: relevant evidence may
occur at the beginning, middle, or end of the observation, and the useful
pruning unit depends on the tool type.

\begin{table}[t]
\centering
\scriptsize
\begin{tabular}{lrr}
\toprule
Source & Raw inputs & Released rows \\
\midrule
SWE-derived & 10{,}713 & 9{,}205 \\
Synthetic positives & 2{,}039 & 1{,}697 \\
Synthetic negatives & --- & 575 \\
\midrule
Total & 12{,}752 & 11{,}477 \\
\bottomrule
\end{tabular}
\caption{Source composition of the released benchmark. Synthetic negatives are
constructed by pairing mismatched queries and tool outputs.}
\label{tab:source-stats}
\end{table}

\begin{table}[t]
\centering
\scriptsize
\resizebox{\columnwidth}{!}{%
\begin{tabular}{lr@{\hspace{0.8em}}lr@{\hspace{0.8em}}lr}
\toprule
Tool & Rows & Tool & Rows & Tool & Rows \\
\midrule
\texttt{read\_file} & 3768 & \texttt{ls} & 347 & \texttt{build\_output} & 168 \\
\texttt{grep} & 1330 & \texttt{type\_check} & 317 & \texttt{docker\_build} & 151 \\
\texttt{git\_log} & 720 & \texttt{git\_blame} & 291 & \texttt{make\_cmake} & 133 \\
\texttt{python} & 698 & \texttt{npm\_build} & 230 & \texttt{kubectl} & 123 \\
\texttt{test\_output} & 546 & \texttt{tsc} & 229 & \texttt{cargo\_build} & 122 \\
\texttt{curl} & 493 & \texttt{npm\_install} & 227 & \texttt{go\_build} & 120 \\
\texttt{pip\_install} & 441 & \texttt{coverage} & 198 & \texttt{mvn\_gradle} & 114 \\
\texttt{lint\_output} & 184 & \texttt{docker\_logs} & 198 & \texttt{terraform} & 103 \\
\texttt{mypy\_pyright} & 95 & \texttt{eslint} & 78 & \texttt{git\_diff} & 53 \\
\bottomrule
\end{tabular}%
}
\caption{Tool-family counts in the released benchmark.}
\label{tab:tool-counts}
\end{table}

\begin{table}[t]
\centering
\scriptsize
\begin{tabular}{lrrr}
\toprule
Tool & Rows & Avg. input & Avg. gold \\
\midrule
\texttt{read\_file} & 3768 & 1677 & 84 \\
\texttt{grep} & 1330 & 779 & 19 \\
\texttt{git\_log} & 720 & 161 & 11 \\
\texttt{python} & 698 & 60 & 28 \\
\texttt{curl} & 493 & 723 & 68 \\
\texttt{pip\_install} & 441 & 438 & 79 \\
\texttt{type\_check} & 317 & 3418 & 39 \\
\texttt{tsc} & 229 & 1444 & 56 \\
remaining tools & 3481 & 914 & 51 \\
\bottomrule
\end{tabular}
\caption{Largest tool families in the dataset. Token counts are averages over
released examples.}
\label{tab:tool-stats}
\end{table}

\section{Model and Evaluation}
Our model is Qwen 3.5 2B \citep{qwen2026qwen35} fine-tuned with LoRA
\citep{hu2022lora,dettmers2023qlora}. The model receives the focused query and
raw tool output and is trained to emit the verbatim extracted text wrapped in
\texttt{<relevant\_lines>} tags. We fine-tune for three epochs with maximum
sequence length 20{,}000, per-device batch size 8, gradient accumulation 4,
learning rate $2\times10^{-4}$, warmup ratio 0.05, and weight decay 0.01. For
serving we merge the LoRA adapter into the base model and run the merged
checkpoint with vLLM. Training was performed on a single NVIDIA A100 80GB GPU.

We choose Qwen 3.5 2B because the Qwen 3.5 family offers strong performance on
code, reasoning, and agent benchmarks while providing small-scale variants
suitable for efficient deployment \citep{qwen2026qwen35}. In our setting, the aim is not to maximize
zero-shot reasoning depth with the largest possible generator, but to learn a
narrow supervised extraction policy that can run cheaply inside existing agent
systems. A 2B backbone is therefore a better fit for this use case than a much
larger decoder, while still remaining strong enough to benefit from
task-specific supervision.

We compare against three zero-shot generative baselines --- Qwen 3.5 35B A3B,
Kimi K2, and the unfine-tuned Qwen 3.5 2B base --- and four simple heuristics:
BM25, First-N, Last-N, and Random. The heuristic baselines keep approximately
10\% of the input lines so that they operate at a compression level similar to
the gold annotations.

The matching unit is the \emph{line}. Given a predicted line set $P$ and a gold
line set $G$, we report recall, exact match, compression, and two overlap
scores. In the main text, F$_1$ denotes a tolerant line-matching score in which
a predicted line matches a gold line if fuzzy substring similarity exceeds
0.5, reducing brittleness from minor formatting differences in generative
output. We also report strict exact-text overlap F$_1$ in
Table~\ref{tab:main-results} for completeness. Our analysis focuses on
\textbf{recall} under strong \textbf{compression}, with F$_1$ as the summary
metric. This reflects the task itself: dropping relevant evidence is usually
more harmful than keeping a slightly larger block.

In practice, the model is intended as a lightweight pre-processing step for
agent systems. In our release it is exposed both as a CLI that can consume
piped tool output and as a vLLM-served model, so existing coding agents can
insert pruning before the next reasoning step without changing their core
planning logic.

\section{Results}

\paragraph{Setup.}
All models are evaluated on the held-out 618-example test set. All generative
models receive the same \texttt{<query>/<tool\_output>} prompt format as the
fine-tuned model. The fine-tuned model is served with vLLM; the larger zero-shot
models are accessed via OpenAI-compatible APIs.

\begin{table}[t]
\centering
\scriptsize
\resizebox{\columnwidth}{!}{%
\begin{tabular}{lcccccc}
\toprule
Model & Prec. & Recall & Strict F$_1$ & Exact & F$_1$ & Compression \\
\midrule
\method{}-2B & \textbf{0.80} & \textbf{0.86} & \textbf{0.79} & \textbf{0.49} & \textbf{0.80} & 0.92 \\
Qwen 3.5 35B A3B & 0.74 & 0.75 & 0.70 & 0.39 & 0.73 & 0.92 \\
Kimi K2 & 0.61 & 0.53 & 0.53 & 0.30 & 0.68 & 0.94 \\
Qwen 3.5 2B (base) & 0.42 & 0.53 & 0.41 & 0.19 & 0.55 & 0.82 \\
\midrule
BM25 (10\%) & 0.13 & 0.22 & 0.13 & 0.01 & 0.23 & 0.90 \\
First-N (10\%) & 0.07 & 0.14 & 0.08 & 0.02 & 0.16 & 0.91 \\
Random (10\%) & 0.07 & 0.10 & 0.07 & 0.01 & 0.20 & 0.91 \\
Last-N (10\%) & 0.05 & 0.05 & 0.04 & 0.01 & 0.14 & 0.91 \\
\bottomrule
\end{tabular}
}
\caption{Held-out test results. Compression is the fraction of input removed.
\method{}-2B is LoRA-tuned; other generative models use the same prompt
zero-shot.}
\label{tab:main-results}
\end{table}

\begin{table*}[!t]
\centering
\scriptsize
\setlength{\tabcolsep}{3pt}
\renewcommand{\arraystretch}{1.05}
\begin{tabularx}{\textwidth}{p{2.2cm}p{4.2cm}p{3.2cm}X}
\toprule
\textbf{Pattern} & \textbf{Query / Observation} & \textbf{\method{}-2B} & \textbf{Baseline error} \\
\midrule
Precise selection in structured output &
\texttt{git\_log}, 21 lines. Query asks for the commit relevant to the dimension-order change in \texttt{xr.polyval}. &
Selects the single correct commit entry. &
Qwen 35B chooses a plausible but wrong transpose-related commit; Qwen 2B base over-selects several \texttt{polyval} commits. \\

Failure-block extraction &
\texttt{service log}, 176 lines. Query asks for the TLS handshake failure affecting the health-check request. &
Returns the correct 5-line health-check failure block. &
Qwen 35B selects a later payment-request TLS failure; Kimi K2 keeps only part of the correct block. \\

Correct empty prediction &
\texttt{docker\_logs}, 316 lines. Query asks for a numpy version conflict that is not present. &
Returns empty output. &
Qwen 35B generates explanatory text (``No relevant lines found\ldots''); Qwen 2B base returns unrelated database errors. \\

Adjacent over-selection &
\texttt{build\_output}, 110 lines. Query asks for the Dockerfile syntax error on line 12. &
Finds the Dockerfile error but also includes a nearby Python \texttt{SyntaxError}. &
Qwen 35B misses the Dockerfile error entirely and selects only the Python block. \\
\bottomrule
\end{tabularx}
\caption{Representative held-out examples. The fine-tuned model is strongest on
precise hits, compact failure blocks, and true negatives; its residual errors
are usually adjacent over-selection.}
\label{tab:qual-results}
\end{table*}

\method{}-2B attains the highest recall among all systems while maintaining
92\% compression. It outperforms the 18$\times$ larger Qwen 3.5 35B A3B by
11 recall points and the unfine-tuned 2B base by 33 points. The gain is not
just recall: the fine-tuned model is also the most precise system in the
comparison, which indicates that it has learned a tool-specific extraction
policy rather than generic instruction following.

The zero-shot baselines differ in how they fail. Qwen 35B is the strongest of
them, but still misses benchmark-specific regularities: in repetitive logs or
Git history it often selects a semantically adjacent but incorrect block.
Kimi K2 is the most aggressive compressor, but pays for that with lower
recall, often dropping lines that look boilerplate-like but are actually part
of the gold evidence. The unfine-tuned 2B base under-compresses and produces
noisier extractions. Heuristic baselines perform substantially worse than any
generative model. BM25, which is effective for document retrieval, reaches
only 0.22 recall on tool output. This is a direct consequence of the task:
relevant lines may occur at the beginning, middle, or end of an observation,
and usefulness depends on the query rather than lexical overlap alone.
Figure~\ref{fig:tradeoff} shows the recall--compression trade-off across all
systems.

\begin{center}
\begin{tikzpicture}
\begin{axis}[
  width=\columnwidth,
  height=4.7cm,
  xmin=0.810, xmax=0.955,
  ymin=0.00, ymax=0.95,
  xlabel={Compression (fraction of input removed)},
  ylabel={Recall},
  grid=both,
  grid style={draw=gray!20},
  major grid style={draw=gray!35},
  tick label style={font=\small},
  label style={font=\small},
]
\addplot[only marks, mark=*, mark size=4pt, blue!80!black]
  coordinates {(0.9150,0.8624)};

\addplot[only marks, mark=square*, mark size=4pt, red!70!black]
  coordinates {(0.9177,0.7498)};

\addplot[only marks, mark=triangle*, mark size=4.5pt, orange!80!black]
  coordinates {(0.9425,0.5286)};

\addplot[only marks, mark=diamond*, mark size=4pt, purple!70!black]
  coordinates {(0.8197,0.5299)};

\addplot[only marks, mark=o, mark size=3pt, gray!60]
  coordinates {(0.9036,0.2172) (0.9055,0.1445) (0.9067,0.1009) (0.9130,0.0503)};

\end{axis}
\end{tikzpicture}

\captionof{figure}{Recall--compression trade-off on the held-out test set.
\textcolor{blue!80!black}{$\bullet$} \method{}-2B,\ 
\textcolor{red!70!black}{$\blacksquare$} Qwen 35B,\ 
\textcolor{orange!80!black}{$\blacktriangle$} Kimi K2,\ 
\textcolor{purple!70!black}{$\blacklozenge$} Qwen 2B,\ 
\textcolor{gray!60}{$\circ$}
heuristic baselines.}
\label{fig:tradeoff}
\end{center}

Qualitatively, the fine-tuned model succeeds for three recurring reasons. It
learns to isolate a single relevant hit in structured outputs such as
\texttt{grep} and \texttt{git\_log}; it learns to keep compact failure blocks
in noisy test, build, and service logs; and it is much better than larger
zero-shot models at returning empty output on explicit negatives. On the 59
negative examples in the held-out test set, it returns empty output 80\% of the
time, compared with 7\% for Qwen 35B. Its strongest
remaining failures are usually semantically adjacent selections rather than
wholly irrelevant output: the wrong file from an \texttt{ls} listing, a
related commit from the same module, or a correct failure block with nearby
extra context attached. Table~\ref{tab:qual-results} summarizes representative
held-out examples. Figure~\ref{fig:qualitative} shows the same behavior on a
compact \texttt{kubectl} observation, where the relevant evidence is a
two-line failure block inside a much longer output.

\begin{figure}[t]
\centering
\footnotesize
\textbf{Query:} \emph{Find the block showing the \texttt{OOMKilled} reason and
exit code for the \texttt{analytics-worker} container.}

\smallskip
\begin{minipage}[t]{0.98\columnwidth}
\fbox{%
\parbox{0.96\linewidth}{%
\ttfamily\scriptsize
{\color{gray}23: State: Waiting}\\
{\color{gray}24: Reason: CrashLoopBackOff}\\
{\color{gray}25: Last State: Terminated}\\
\colorbox{yellow!25}{26: Reason: OOMKilled}\\
\colorbox{yellow!25}{27: Exit Code: 137}\\
{\color{gray}28: Started: Mon, 10 Mar 2026 08:12:45 +0000}\\
{\color{gray}\quad\ldots}\\
{\color{gray}30: Ready: False}\\
{\color{gray}31: Restart Count: 3}
}}
\end{minipage}

\smallskip
\begin{minipage}[t]{0.98\columnwidth}
\textbf{Gold span}

\vspace{2pt}
\fbox{%
\parbox{0.72\linewidth}{%
\ttfamily\scriptsize
26: Reason: OOMKilled\\
27: Exit Code: 137
}}
\end{minipage}
\caption{Compact qualitative example from a \texttt{kubectl} observation. The
full output has 250 lines; the gold evidence has two.}
\label{fig:qualitative}
\end{figure}

\section{Limitations}
The benchmark evaluates pruning quality on single tool observations rather than
full agent trajectories. It therefore measures evidence preservation directly,
but not the downstream effect on end-to-end task completion. In addition,
usefulness is approximated with span overlap, which cannot capture every valid
alternative pruning decision. Finally, some tool families remain noisier than
others, especially \texttt{grep} and \texttt{lint\_output}.

\section{Conclusion}
\method{} is a task-conditioned tool-output pruner for coding agents. The
benchmark pairs grounded span annotations over 27 tool types with real
SWE-bench workflows and synthetic multi-ecosystem observations. A compact
LoRA-tuned Qwen 3.5 2B model reaches 0.86 recall while removing 92\% of
input tokens, outperforming the 18$\times$ larger Qwen 3.5 35B A3B and all
heuristic baselines by a wide margin. The broader lesson is that mixed-format
tool output is not handled well by zero-shot generative models or retrieval
heuristics alone; it responds well to narrow, task-specific supervision.

The model and evaluation code are available on
\href{https://github.com/KRLabsOrg/squeez}{GitHub}, and the released model and
dataset are available on
\href{https://huggingface.co/KRLabsOrg/squeez-2b}{Hugging Face}.

\bibliography{references}

@misc{hu2022lora,
      title={LoRA: Low-Rank Adaptation of Large Language Models}, 
      author={Edward J. Hu and Yelong Shen and Phillip Wallis and Zeyuan Allen-Zhu and Yuanzhi Li and Shean Wang and Lu Wang and Weizhu Chen},
      year={2021},
      eprint={2106.09685},
      archivePrefix={arXiv},
      primaryClass={cs.CL},
      url={https://arxiv.org/abs/2106.09685}, 
}

@inproceedings{dettmers2023qlora,
 author = {Dettmers, Tim and Pagnoni, Artidoro and Holtzman, Ari and Zettlemoyer, Luke},
 booktitle = {Advances in Neural Information Processing Systems},
 editor = {A. Oh and T. Naumann and A. Globerson and K. Saenko and M. Hardt and S. Levine},
 pages = {10088--10115},
 publisher = {Curran Associates, Inc.},
 title = {QLoRA: Efficient Finetuning of Quantized LLMs},
 url = {https://proceedings.neurips.cc/paper_files/paper/2023/file/1feb87871436031bdc0f2beaa62a049b-Paper-Conference.pdf},
 volume = {36},
 year = {2023}
}

@inproceedings{jiang2023llmlingua,
    title = "{LLML}ingua: Compressing Prompts for Accelerated Inference of Large Language Models",
    author = "Jiang, Huiqiang  and
      Wu, Qianhui  and
      Lin, Chin-Yew  and
      Yang, Yuqing  and
      Qiu, Lili",
    editor = "Bouamor, Houda  and
      Pino, Juan  and
      Bali, Kalika",
    booktitle = "Proceedings of the 2023 Conference on Empirical Methods in Natural Language Processing",
    month = dec,
    year = "2023",
    address = "Singapore",
    publisher = "Association for Computational Linguistics",
    url = "https://aclanthology.org/2023.emnlp-main.825/",
    doi = "10.18653/v1/2023.emnlp-main.825",
    pages = "13358--13376"
}

@inproceedings{jiang2024longllmlingua,
    title = "{L}ong{LLML}ingua: Accelerating and Enhancing {LLM}s in Long Context Scenarios via Prompt Compression",
    author = "Jiang, Huiqiang  and
      Wu, Qianhui  and
      Luo, Xufang  and
      Li, Dongsheng  and
      Lin, Chin-Yew  and
      Yang, Yuqing  and
      Qiu, Lili",
    editor = "Ku, Lun-Wei  and
      Martins, Andre  and
      Srikumar, Vivek",
    booktitle = "Proceedings of the 62nd Annual Meeting of the Association for Computational Linguistics (Volume 1: Long Papers)",
    month = aug,
    year = "2024",
    address = "Bangkok, Thailand",
    publisher = "Association for Computational Linguistics",
    url = "https://aclanthology.org/2024.acl-long.91/",
    doi = "10.18653/v1/2024.acl-long.91",
    pages = "1658--1677"
}

@inproceedings{hwang2024exit,
    title = "{EXIT}: Context-Aware Extractive Compression for Enhancing Retrieval-Augmented Generation",
    author = "Hwang, Taeho  and
      Cho, Sukmin  and
      Jeong, Soyeong  and
      Song, Hoyun  and
      Han, SeungYoon  and
      Park, Jong C.",
    editor = "Che, Wanxiang  and
      Nabende, Joyce  and
      Shutova, Ekaterina  and
      Pilehvar, Mohammad Taher",
    booktitle = "Findings of the Association for Computational Linguistics: ACL 2025",
    month = jul,
    year = "2025",
    address = "Vienna, Austria",
    publisher = "Association for Computational Linguistics",
    url = "https://aclanthology.org/2025.findings-acl.253/",
    doi = "10.18653/v1/2025.findings-acl.253",
    pages = "4895--4924",
    ISBN = "979-8-89176-256-5"
}

@misc{chirkova2025provence,
      title={Provence: efficient and robust context pruning for retrieval-augmented generation}, 
      author={Nadezhda Chirkova and Thibault Formal and Vassilina Nikoulina and Stéphane Clinchant},
      year={2025},
      eprint={2501.16214},
      archivePrefix={arXiv},
      primaryClass={cs.CL},
      url={https://arxiv.org/abs/2501.16214}, 
}

@misc{ayanami2026swepruner,
      title={SWE-Pruner: Self-Adaptive Context Pruning for Coding Agents}, 
      author={Yuhang Wang and Yuling Shi and Mo Yang and Rongrui Zhang and Shilin He and Heng Lian and Yuting Chen and Siyu Ye and Kai Cai and Xiaodong Gu},
      year={2026},
      eprint={2601.16746},
      archivePrefix={arXiv},
      primaryClass={cs.SE},
      url={https://arxiv.org/abs/2601.16746},
}

@article{focusagent2025,
      title={FocusAgent: Simple Yet Effective Ways of Trimming the Large Context of Web Agents}, 
      author={Imene Kerboua and Sahar Omidi Shayegan and Megh Thakkar and Xing Han Lù and Léo Boisvert and Massimo Caccia and Jérémy Espinas and Alexandre Aussem and Véronique Eglin and Alexandre Lacoste},
      year={2025},
      eprint={2510.03204},
      archivePrefix={arXiv},
      primaryClass={cs.CL},
      url={https://arxiv.org/abs/2510.03204}, 
}

@inproceedings{kovacs2025verbatim,
    title = "{KR} Labs at {A}rch{EHR}-{QA} 2025: A Verbatim Approach for Evidence-Based Question Answering",
    author = "Kovacs, Adam and Schmitt, Paul and Recski, Gabor",
    editor = "Soni, Sarvesh and Demner-Fushman, Dina",
    booktitle = "Proceedings of the 24th Workshop on Biomedical Language Processing (Shared Tasks)",
    month = aug,
    year = "2025",
    address = "Vienna, Austria",
    publisher = "Association for Computational Linguistics",
    url = "https://aclanthology.org/2025.bionlp-share.8/",
    doi = "10.18653/v1/2025.bionlp-share.8",
    pages = "69--74"
}

@misc{zilliz2025semantic,
  title={Semantic Highlight Bilingual Model},
  author={{Zilliz}},
  year={2025},
  howpublished={\url{https://huggingface.co/zilliz/semantic-highlight-bilingual-v1}},
  note={Model card}
}

@inproceedings{rajpurkar2016squad,
    title = "{SQ}u{AD}: 100,000+ Questions for Machine Comprehension of Text",
    author = "Rajpurkar, Pranav  and
      Zhang, Jian  and
      Lopyrev, Konstantin  and
      Liang, Percy",
    editor = "Su, Jian  and
      Duh, Kevin  and
      Carreras, Xavier",
    booktitle = "Proceedings of the 2016 Conference on Empirical Methods in Natural Language Processing",
    month = nov,
    year = "2016",
    address = "Austin, Texas",
    publisher = "Association for Computational Linguistics",
    url = "https://aclanthology.org/D16-1264/",
    doi = "10.18653/v1/D16-1264",
    pages = "2383--2392"
}

@inproceedings{thorne2018fever,
    title = "{FEVER}: a Large-scale Dataset for Fact Extraction and {VER}ification",
    author = "Thorne, James  and
      Vlachos, Andreas  and
      Christodoulopoulos, Christos  and
      Mittal, Arpit",
    editor = "Walker, Marilyn  and
      Ji, Heng  and
      Stent, Amanda",
    booktitle = "Proceedings of the 2018 Conference of the North {A}merican Chapter of the Association for Computational Linguistics: Human Language Technologies, Volume 1 (Long Papers)",
    month = jun,
    year = "2018",
    address = "New Orleans, Louisiana",
    publisher = "Association for Computational Linguistics",
    url = "https://aclanthology.org/N18-1074/",
    doi = "10.18653/v1/N18-1074",
    pages = "809--819"
}

@inproceedings{yang2018hotpotqa,
    title = "{H}otpot{QA}: A Dataset for Diverse, Explainable Multi-hop Question Answering",
    author = "Yang, Zhilin  and
      Qi, Peng  and
      Zhang, Saizheng  and
      Bengio, Yoshua  and
      Cohen, William  and
      Salakhutdinov, Ruslan  and
      Manning, Christopher D.",
    editor = "Riloff, Ellen  and
      Chiang, David  and
      Hockenmaier, Julia  and
      Tsujii, Jun{'}ichi",
    booktitle = "Proceedings of the 2018 Conference on Empirical Methods in Natural Language Processing",
    month = oct # "-" # nov,
    year = "2018",
    address = "Brussels, Belgium",
    publisher = "Association for Computational Linguistics",
    url = "https://aclanthology.org/D18-1259/",
    doi = "10.18653/v1/D18-1259",
    pages = "2369--2380"
}

@inproceedings{see2017get,
    title = "Get To The Point: Summarization with Pointer-Generator Networks",
    author = "See, Abigail  and
      Liu, Peter J.  and
      Manning, Christopher D.",
    editor = "Barzilay, Regina  and
      Kan, Min-Yen",
    booktitle = "Proceedings of the 55th Annual Meeting of the Association for Computational Linguistics (Volume 1: Long Papers)",
    month = jul,
    year = "2017",
    address = "Vancouver, Canada",
    publisher = "Association for Computational Linguistics",
    url = "https://aclanthology.org/P17-1099/",
    doi = "10.18653/v1/P17-1099",
    pages = "1073--1083"
}

@inproceedings{liu2019text,
    title = "Text Summarization with Pretrained Encoders",
    author = "Liu, Yang  and
      Lapata, Mirella",
    editor = "Inui, Kentaro  and
      Jiang, Jing  and
      Ng, Vincent  and
      Wan, Xiaojun",
    booktitle = "Proceedings of the 2019 Conference on Empirical Methods in Natural Language Processing and the 9th International Joint Conference on Natural Language Processing (EMNLP-IJCNLP)",
    month = nov,
    year = "2019",
    address = "Hong Kong, China",
    publisher = "Association for Computational Linguistics",
    url = "https://aclanthology.org/D19-1387/",
    doi = "10.18653/v1/D19-1387",
    pages = "3730--3740"
}

@misc{qwen2026qwen35,
  title={Qwen3.5: Towards Native Multimodal Agents},
  author={{Qwen Team}},
  year={2026},
  howpublished={\url{https://qwen.ai/blog?id=qwen3.5}},
  note={Blog post}
}

@inproceedings{yang2024sweagent,
author = {Yang, John and Jimenez, Carlos E. and Wettig, Alexander and Lieret, Kilian and Yao, Shunyu and Narasimhan, Karthik and Press, Ofir},
title = {SWE-agent: agent-computer interfaces enable automated software engineering},
year = {2024},
isbn = {9798331314385},
publisher = {Curran Associates Inc.},
address = {Red Hook, NY, USA},
booktitle = {Proceedings of the 38th International Conference on Neural Information Processing Systems},
articleno = {1601},
numpages = {125},
location = {Vancouver, BC, Canada},
series = {NIPS '24}
}

@misc{wang2024opendevin,
      title={OpenHands: An Open Platform for AI Software Developers as Generalist Agents}, 
      author={Xingyao Wang and Boxuan Li and Yufan Song and Frank F. Xu and Xiangru Tang and Mingchen Zhuge and Jiayi Pan and Yueqi Song and Bowen Li and Jaskirat Singh and Hoang H. Tran and Fuqiang Li and Ren Ma and Mingzhang Zheng and Bill Qian and Yanjun Shao and Niklas Muennighoff and Yizhe Zhang and Binyuan Hui and Junyang Lin and Robert Brennan and Hao Peng and Heng Ji and Graham Neubig},
      year={2025},
      eprint={2407.16741},
      archivePrefix={arXiv},
      primaryClass={cs.SE},
      url={https://arxiv.org/abs/2407.16741}, 
}

@inproceedings{jimenez2024swebench,
    title={{SWE}-bench: Can Language Models Resolve Real-world Github Issues?},
    author={Carlos E Jimenez and John Yang and Alexander Wettig and Shunyu Yao and Kexin Pei and Ofir Press and Karthik R Narasimhan},
    booktitle={The Twelfth International Conference on Learning Representations},
    year={2024},
    url={https://openreview.net/forum?id=VTF8yNQM66}
}

\end{document}